\documentclass[a4paper,12pt]{article}
\pdfoutput=1
\usepackage{feynmp-auto,expdlist}
\usepackage{amsmath, amsfonts, amssymb,mathbbol}
\usepackage{graphicx}
\usepackage{enumerate,cancel}
\usepackage{hyperref}
\usepackage{latexsym}
\usepackage{dsfont}
\usepackage{hepnicenames}
\usepackage{enumerate}
\usepackage{soul}
\usepackage[normalem]{ulem}
\usepackage{comment}

\usepackage{units}

\newcommand{\gfootnote}[1]{\footnote{\tiny\color{gray}#1}}
\renewcommand{\gfootnote}[1]{}
\usepackage{soul}

\usepackage{mathrsfs,graphicx,rotating,amsmath,amsfonts,mathtools,booktabs,amssymb,wasysym}
\usepackage{hyperref}\usepackage{slashed}
\usepackage[nosort]{cite}
\usepackage[table,xcdraw,dvipsnames]{xcolor}
\usepackage{bm}
\usepackage{graphicx}
\usepackage{multirow,multicol}
\hypersetup{colorlinks,bookmarksopen,bookmarksnumbered,
	linkcolor=blus,pdfstartview=FitH,urlcolor=rossos,citecolor=verde}
\allowdisplaybreaks

\renewcommand{\[}{\left[}

\newcommand{\Lag}{\mathscr{L}}

\newcommand{\mio}[1]{}

\newcommand{\med}[1]{\langle #1\rangle}

\newcommand{\bpm}{\begin{pmatrix}}
\newcommand{\epm}{\end{pmatrix}}

\usepackage{mathrsfs}

\newcommand{\fig}[1]{~\ref{fig:#1}}

\allowdisplaybreaks
\usepackage{multicol}
\usepackage{color}
\definecolor{rosso}{cmyk}{0,1,1,0.4}
\definecolor{rossos}{cmyk}{0,1,1,0.55}
\definecolor{rossoc}{cmyk}{0,1,1,0.2}
\definecolor{blu}{cmyk}{1,1,0,0.3}
\definecolor{blus}{cmyk}{1,1,0,0.6}
\definecolor{bluc}{cmyk}{1,1,0,0.1}
\definecolor{verde}{cmyk}{0.92,0,0.59,0.25}
\definecolor{verdec}{cmyk}{0.92,0,0.59,0.15}
\definecolor{verdes}{cmyk}{0.92,0,0.59,0.4}

\newcommand{\eq}[1]{~{\rm (\ref{eq:#1})}}

\newcommand{\beq}{\begin{equation}}
\newcommand{\eeq}{\end{equation}}

\newcommand{\bea}{\begin{eqnarray}}
\newcommand{\eea}{\end{eqnarray}}
\newcommand{\be}{\begin{equation}}
\newcommand{\ee}{\end{equation}}
\font\tenrsfs=rsfs10 at 12pt
\font\sevenrsfs=rsfs7
\font\fiversfs=rsfs5
\newfam\rsfsfam
\textfont\rsfsfam=\tenrsfs
\scriptfont\rsfsfam=\sevenrsfs
\scriptscriptfont\rsfsfam=\fiversfs
\def\be#1\ee{\begin{equation}#1\end{equation}}
\def\bl#1\el{\begin{align}#1\end{align}}
\def\ba#1\ea{\begin{align*}#1\end{align*}}

\renewenvironment{thebibliography}[1]
{\begin{multicols}{2}[\section*{\refname}]%
		\@mkboth{\MakeUppercase\refname}{\MakeUppercase\refname}%
		\list{\@biblabel{\@arabic\c@enumiv}}%
		{\settowidth\labelwidth{\@biblabel{#1}}%
			\leftmargin\labelwidth
			\advance\leftmargin\labelsep
			\@openbib@code
			\usecounter{enumiv}%
			\let\p@enumiv\@empty
			\renewcommand\theenumiv{\@arabic\c@enumiv}}%
		\sloppy
		\clubpenalty4000
		\@clubpenalty \clubpenalty
		\widowpenalty4000%
		\sfcode`\.\@m}
	{\def\@noitemerr
		{\@latex@warning{Empty `thebibliography' environment}}%
		\endlist\end{multicols}}

\makeatletter
\font\ital=cmu10

\def\hhref#1{\href{http://arxiv.org/abs/#1}{arXiv:#1}}
\usepackage{xstring}
\newcommand{\hhrefq}[1]{\IfSubStr{#1}{:}{\href{http://inspirehep.net/search?ln=en&ln=en&p=#1&of=hb&action_search=Search&sf=&so=d&rm=&rg=25&sc=0}{InSpire:#1}}{\hhref{#1}}}

\def\art{\@ifnextchar[{\eart}{\oart}}
\def\eart[#1]#2#3#4#5#6{{\rm #2}, {\em #3 \bf #4} {\rm (#6) #5} ({\em #1})}
\def\article{\@ifnextchar[{\earticle}{\oarticle}}
\def\oarticle#1#2#3#4#5#6{{\rm #1}, {\ital `#6'}, {\rm #2 #3 (#5) #4}}
\def\earticle[#1]#2#3#4#5#6#7{{\rm #2}, {\ital `#7'}, {\rm #3 #4 (#6) #5}  [\hhrefq{#1}]}
\def\hepart[#1]#2{{\rm #2, \sl#1}}
\def\heparticle[#1]#2#3{#2, {\ital `#3'} [\hhrefq{#1}]}
\newcommand{\doi}[1]{\href{http://dx.doi.org/#1}{[link]}}

\newcommand{\hhrefqq}[1]{\IfBeginWith{#1}{10.}{\href{https://doi.org/#1}{doi:#1}}{\hhrefq{#1}}}

\renewenvironment{thebibliography}[1]
{\begin{multicols}{2}[\section*{\refname}]%
		\@mkboth{\MakeUppercase\refname}{\MakeUppercase\refname}%
		\list{\@biblabel{\@arabic\c@enumiv}}%
		{\settowidth\labelwidth{\@biblabel{#1}}%
			\leftmargin\labelwidth
			\advance\leftmargin\labelsep
			\@openbib@code
			\usecounter{enumiv}%
			\let\p@enumiv\@empty
			\renewcommand\theenumiv{\@arabic\c@enumiv}}%
		\sloppy
		\clubpenalty4000
		\@clubpenalty \clubpenalty
		\widowpenalty4000%
		\sfcode`\.\@m}
	{\renewcommand{\@noitemerr}
		{\@latex@warning{Empty `thebibliography' environment}}%
		\endlist\end{multicols}}

\newcommand{\eqnsystem}[1]{
	\renewcommand{\@eqnnum}{{\rm (\thealphaequation)}}
	\renewcommand{\@@eqncr}{\let\@tempa\relax \ifcase\@eqcnt \def\@tempa{& & &} \or
		\newcommand{\@tempa}{& &}\or \newcommand{\@tempa}{&}\fi\@tempa
		\if@eqnsw\@eqnnum\refstepcounter{alphaequation}\fi
		\global\@eqnswtrue\global\@eqcnt=0\cr}
	\refstepcounter{equation} \let\@currentlabel\theequation \def\@tempb{#1}
	\ifx\@tempb\empty\else\label{#1}\fi
	\refstepcounter{alphaequation}
	\let\@currentlabel\thealphaequation
	\global\@eqnswtrue\global\@eqcnt=0 \tabskip\@centering\let\\=\@eqncr
	$$\halign to \displaywidth\bgroup \@eqnsel\hskip\@centering
	$\displaystyle\tabskip\z@{##}$&\global\@eqcnt\@ne
	\hskip2\arraycolsep\hfil${##}$\hfil& \global\@eqcnt\tw@\hskip2\arraycolsep
	$\displaystyle\tabskip\z@{##}$\hfil
	\tabskip\@centering&\llap{##}\tabskip\z@\cr}
\def\endeqnsystem{\@@eqncr\egroup$$\global\@ignoretrue} \makeatother

\newcounter{alphaequation}[equation]
\renewcommand{\thealphaequation}{\theequation\hbox to
	0.6em{\hfil\alph{alphaequation}\hfil}}

\definecolor{Gray}{gray}{0.95}

\begin{document}

\begin{center}
\bigskip\bigskip
{\bf\LARGE\color{red!60!black} Testing the arrow of time at the cosmo collider}\\
\bigskip\bigskip
{\bf Shuntaro Aoki}$^a$ and {\bf Alessandro Strumia}$^b$  \\[2ex]
{\it $^a$ RIKEN Center for Interdisciplinary Theoretical and Mathematical Sciences, Saitama, Japan\\[1ex]
$^b$ Dipartimento di Fisica, Universit\`a di Pisa, Italia}\\
\bigskip
{\bf\large\color{blus} Abstract}
\begin{quote}\large
Normal particles carry a microscopic arrow of causality. 
Lee–Wick ghosts carry the reversed arrow, mediating characteristic collider signals in flat space: 
opposite-sign scattering amplitudes that violate positivity bounds;
acausality on time scales set by their negative decay rate.
During inflation, the corresponding cosmo-collider ghost signals are:
opposite-sign non-Gaussianities; 
Boltzmann-unsuppressed local oscillatory signals without their non-local counterparts;
IR-enhanced bi-spectrum and power spectrum, depending on the dimension of the interaction operator, which 
decreases if the ghost decay rate is comparable to the Hubble rate.
\end{quote}
\end{center}
\thispagestyle{empty}

\tableofcontents

\newpage

\section{Introduction}
Cosmology provides a window into particle physics at energy scales far beyond terrestrial collider experiments.
The early universe, during the epoch of cosmic inflation, acted as a high-energy collider
that might have produced heavy particles with masses of order the inflationary Hubble scale, $M \sim H$,
that left characteristic oscillatory features  in the 
correlation functions of curvature perturbations~\cite{0911.3380,1109.0292, 1211.1624,1503.08043}.
The frequency of oscillations as function of small ratios of momenta is set by the mass of the exchanged particle. 
The detailed structure for comparable momenta carries extra information~\cite{2205.01692}.
Many studies analysed a variety of models, e.g.~\cite{1610.06597,1612.08122,1805.02656,1811.11200,1907.07390,2004.02887,2112.10793,2203.06349,2408.07069}, 
elucidating how the spin\cite{1503.08043,1607.03735}, interactions~\cite{2205.01692}, and mass spectrum of heavy particles~\cite{1610.06597,1612.08122,2012.13667,2112.05710,2309.10849,2404.09547} are imprinted in primordial correlators. 

\smallskip

However, in simplest scenarios, the amplitude of these signals is expected to be too small to be detected by upcoming surveys. 
Nevertheless, technically challenging computations are being performed 
hoping  that nature might have been generous, through couplings as large as possible, or via non-minimal sources of inhomogeneities beyond the inflaton.
Tree level exchange of some extra scalar  emerges as the simplest plausible possibility for a detectable signal~\cite{0911.3380},
as higher spins are loops giving more suppressed effects.

We here focus on tree-level scalar exchange, and explore how
a cosmo-collider signal would probe an interesting unusual feature of the extra scalar: its {\em arrow of time}.

Ordinary particles carry an intrinsic arrow of time. This fact is missed by some authors (even of books), so we summarise it.
This arrow arises because all known particles have positive kinetic energy such that their quantum Hamiltonian $H$ has a vacuum state with minimal energy, and 
time evolution $e^{-i H t}$ flows in one direction~\cite{1908.04170,2003.09047}.
More technically, the T and CPT symmetries are anti-unitary so that quantum mechanics is time-covariant, but not time-symmetric~\cite{2003.09047}.
For example, the cause precedes the effect when two particles scatter producing a resonance, that decays after having being produced.
The thermodynamic arrow of time follows from this microscopic arrow of causality.
So one remembers the past and ages in the future.
The universe expansion sets a different arrow of time, that negligibly affects ordinary particle processes.

General  causality bounds on cosmo-collider signals have been explored in~\cite{2502.04215,2506.19198}.
The interesting concrete example of new physics that could violate such bounds are  ghost particles with their opposite arrow of time.
Taken in isolation, ghost particles are as well behaved as normal particles: they are local but anti-causal.
The ghost action is similar to the action of normal particles, up to an overall minus sign,  implying a negative classical kinetic energy.
Possibly problematic new physics effects arise in theories where normal particles interact with ghosts,
leading to acausality on time scales comparable to the ghost decay rate, as
studied by Lee, Wick and others~\cite{0805.2156}.

In section~\ref{sec:flat} we summarise ghosts in flat space, showing how they affect correlators similar to the ones
testable at the cosmo collider.
In section~\ref{sec:dS} we try making sense of ghosts in a nearly de Sitter background, two subtle topics.
Results are used to discuss  how the opposite arrow of time of ghosts would appear in cosmo-collider signals.
Conclusions are given in section~\ref{concl}.

%

\section{What is a ghost in flat space?}\label{sec:flat}
In this section we summarise attempts to make sense of a ghost in flat Minkowski space,
before addressing the inflationary de Sitter background in the next section.

\subsection{Normal scalar in flat space}
To highlight the differences with a ghost, we start briefly recalling basic aspects of a `normal' scalar field $\phi(x)$ with positive kinetic energy in its free action 
$S_+ = \frac12 \int d^4x\, [ (\partial_\mu\phi)^2 - m^2 \phi^2]$.
The field is expanded in modes as
\beq \label{eq:normalphi}
\phi(x) =\int \frac{d^3k}{(2\pi)^32E_k} [a_k e^{-ik\cdot x}+b_k^\dagger e^{ik\cdot x}],\qquad
\eeq
such that the desired commutator $[\phi(x),\pi(y)]=i\delta(\vec x - \vec y)$ with $\pi = \dot \phi^\dagger$ 
is obtained for
$[a_k, a_{k'}^\dagger]=2E_k(2\pi)^3\delta(\vec k-\vec k')$.
If $\phi$ is real the operators satisfy $b=a$; if $\phi$ is complex similar commutation relations hold for $b$. 
The time evolution operator is given by the time-ordered expression
$U = {\rm  T}\, e^{i \int H_{\rm int} dt}$ and path integrals compute averages of time-ordered fields.
So perturbation theory employs the time-ordered Feynman propagator, 
$\langle  0 | {\rm T}\, \phi(x)\phi^\dagger (y)|0\rangle$, that propagates positive energy forward in time.
It is given by
\beq  \langle  0 | \theta(x_0-y_0) \phi(x)\phi^\dagger(y)+\theta(y_0-x_0)\phi^\dagger(y)\phi(x) |0\rangle  =
i \int\frac{d^4k}{(2\pi)^4} \frac{e^{-ik\cdot (x-y)}}{k^2-m^2+i\epsilon}\eeq
where the $i \epsilon$ implies a positive spectral density $+\delta (k^2-m^2)$.
The path integral $\int D\phi\,e^{i S_+ }$ is well defined thanks to the small imaginary mass, $S_+= \frac12 \int d^4k \, \phi(k) (k^2 - m^2 + i \epsilon) \phi(k)$,
where $\phi(k) \equiv (2\pi)^{-2} \int d^4 x \, e^{i k\cdot x} \phi(x)$.
The propagator poles allow the continuation to Euclidean time as  $ t_E =+ i t$.
A decaying scalar acquires a positive decay width $\Gamma>0$, such that its resummed propagator becomes $i/(k^2 - m^2 + i m \Gamma)$, in agreement with the $i \epsilon$ prescription.

\subsection{Ghost in flat space}

Next, we consider a scalar `ghost'  with negative kinetic term in its free action, $S_- = -  S_+$.
The ghost action can be quantised as: 
i) negative energy and positive norm, or as
ii) positive energy and undefined norm~\cite{0805.2156}. 
We discard the first problematic possibility, and attempt the second possibility.
This quantization means that $\phi$ remains as in eq.\eq{normalphi}, but now  $\pi = - \dot\phi^\dagger$.
So the desired $[\phi,\pi]$ commutator is obtained for opposite sign $[a_k, a_{k'}^\dagger]$, implying undefined norm.
As a result a ghost has opposite-sign propagator $-1/(k^2 - m^2+ i \epsilon)$ with $\epsilon>0$.
Having the same sign of $i \epsilon$ as for a normal particle, the standard analytic continuation to the Euclidean $ t_E = i t$ remains valid, 
preserving standard power-counting renormalizability.
At this stage, the extra minus sign `only' implies a negative ghost spectral density and
ghost-mediated amplitudes that violate positivity bounds.

\begin{figure}[t]
$$\includegraphics[width=0.8\textwidth]{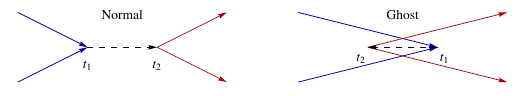}$$
\begin{center}
\caption{\em Resonant $s$-channel production of a normal particle with decay width $\Gamma>0$ (left) and of a ghost with $\Gamma<0$ (right).
\label{fig:acausal}}
\end{center}
\end{figure}

Various authors tried to make sense of this quantization. 
The problem is that negative norms naively imply negative probabilities.
Lee and Wick suggested a partial way around the issue:
since experiments indicate that ghosts, if they exist at all, would appear only as heavy and unstable states, 
one can restrict attention to their observable effects in scattering processes among normal particles with positive norm.
Even so, new (possibly pathological) physics arises in theories where ghosts interact with normal particles.

\smallskip

We already mentioned that virtual exchange of one ghost mediates opposite-sign amplitudes.
More interestingly, loop corrections to the ghost propagator
contain standard imaginary parts, meaning that the ghost acquires a negative decay width $\Gamma<0$.
Consequently, scattering processes among ordinary particles can produce an on-shell ghost that {\em decays  before being produced}~\cite{0805.2156}.
As illustrated in fig.\fig{acausal}, measurable decay vertices are displaced in the direction opposite to the momenta,
signalling acausality on short time-scales comparable to the ghost life-time.
In broader terms, this means that the ghost acquires an arrow of time opposite to that of ordinary particles.
Indeed the ghost resummed propagator is $-i/(k^2 - m^2 + i m \Gamma)$.
Taking the negative width into account, the path integral with action $S_-$
is equivalent to the positive energy scalar with time reversed, $\int D\phi\,e^{i S_- } =\int D\phi \, e^{-i S_+ }$.
The two theories are dual, related by $t \to - t$, which corresponds to an opposite arrow of time.

\medskip

The difficulty is that the finite width $\Gamma$ appears to clash with the assumed $i\epsilon$ prescription.
A common attempt to sidestep this conflict is to assume that the analytic continuation to Euclidean space remains valid,
and to define energy integrals along a contour in the complex plane that deviates from the real axis so as to pass above the acausal pole~\cite{0805.2156}.
Ghosts might be problematic at loop level. 
See~\cite{1512.01237,1709.04925,1806.03605,1908.02416,2004.00376,2110.02246} for recent attempts of making full sense of ghosts, 
mostly motivated by renormalizable quadratic gravity~\cite{Stelle:1976gc}.
We will compute tree-level effects.

\medskip


\subsection{Computing amplitudes and correlators}
We are interested in cosmo-collider effects of the arrow of time analogous to collider decay vertices displaced in the acausal way, as in fig.\fig{acausal}.
Scattering amplitudes at colliders are conveniently computed using the usual $(E_k, \vec{k})$ variables, as both energy and momentum are conserved.
Cosmo-collider effects are usually computed using the $(t,\vec{k})$ variables, 
since the time-dependent cosmological background breaks energy conservation. 
Furthermore, the cosmo-collider observables are field correlators rather than scattering amplitudes
as in flat-space collider physics.
These correlators are computed using the Schwinger–Keldysh (SK) formalism.

\smallskip

We summarize the formalism for a normal scalar in flat space~\cite{1703.10166}.
The Schwinger–Keldysh formalism computes averages of operators ${\cal O}$ at time $t$,
given an initial state at time $t_0$ described by a density matrix $\rho(t_0)$.
This is later specialised to the vacuum at $t_0 \to - \infty$.
So $\med{{\cal O}(t)} = {\rm Tr}\,[ \rho(t_0) U(t_0 \to t) {\cal O}(t) U(t\to t_0)]$.
The density matrix needs to be evolved forward and backward in time to get the desired quantity.
Time-dependent perturbation theory is extended to quantum field theory by computing via a 
path integral with doubled field content.
For example one scalar $\phi$ is doubled to two scalars $\phi_\pm$, as
\beq \label{eq:SKZ}
Z[J_+, J_-]=\int D\phi_+D\phi_- e^{i S[\phi_+] - i S[\phi_-] + \int d^4x( J_+ \phi_+ +J_- \phi_-)}.\eeq
with opposite sign $k\tau\to (1\mp i\epsilon)k\tau$ deformations of the mode functions for $\phi_\pm$,
such that $\phi_-$ evolves backwards in time~\cite{1703.10166}.
So the SK formalism employs four propagators.
Writing the modes as
$ e^{-i k \cdot x}/\sqrt{2E_k}= u_k(t) \exp( i \vec{k}\cdot\vec{x})$, the four propagators in flat space are
\begin{align} 
D_{-+}(k; t_1, t_2) &=u_k(t_1) u_k^*(t_2)=\frac{e^{- i E_k (t_1-t_2)}}{2E_k} = D^*_{+-}(k; t_1, t_2)  ,
\nonumber \\
D_{+-} (k; t_1, t_2) &=u_k^*(t_1) u_k(t_2)=\frac{e^{+ i E_k (t_1-t_2)}}{2E_k} = D^*_{-+}(k; t_1, t_2),   \label{eq:D++}
\\
D_{\pm\pm}(k;t_1,t_2)&=  \theta(t_1-t_2) D_{-+} (k;t_1,t_2)+\theta(t_2-t_1) D_{+-}(k;t_1,t_2), 
\nonumber
\end{align}
with $E_k=+\sqrt{\vec{k}^2 + m^2}$.
$D_{++}$ is the usual  time-ordered Feynman propagator; 
$D_{--} = D_{++}^*$  is its anti-time-ordered correspondent
(with no extra minus sign as,  in the conventional notation, the sign arising from the conjugation
is moved from propagators to vertices~\cite{1703.10166}).


For illustration we compute, using the SK formalism, a theory featuring a massless scalar $\zeta$ and a massive $\phi$ with a cubic interaction $\zeta^2 \phi$.
\begin{itemize}
\item The usual $s$-channel scattering amplitude mediated by a $\phi$ particle with momentum $\vec{k}_s=\vec{k}_1 + \vec{k}_2=\vec{k}_3 + \vec{k}_4$
produced at $x_1$ and absorbed at $x_2$ arises as~\cite{2308.14802}
\beq A (\zeta_{k_1}\zeta_{k_2} \to \phi_{k_s}\to \zeta_{k_3} \zeta_{k_4}) \sim \int_{-\infty}^{+\infty} dt_1 dt_2\, e^{i k_{12} t_1- i k_{34} t_2} D_{++}(k_s; t_1, t_2)
\sim -\frac{\delta(k_{12}-k_{34})}{k_{12}^2-E_{k_s}^2 + i \epsilon}
\eeq
where we omitted couplings and defined $k_i = |\vec{k}_i|$, $k_{12}\equiv k_1+k_2$, $k_{34}=k_3+k_4$.
The denominator is  the usual $s-m^2 + i \epsilon$, that gets resonant on-shell.
\item The contribution to the $\zeta$ field correlator at $t=0$ mediated by a $\phi$ scalar in the $s$-channel is~\cite{2308.14802}
\beq \label{eq:flatcor}\hspace{-4ex}
\med{\zeta_{k_1} \zeta_{k_2} \zeta_{k_3} \zeta_{k_4}} \sim \sum_{ab} ab  \int_{-\infty}^0 dt_1 \, dt_2 \, e^{i a k_{12} t_1+i b k_{34} t_2} D_{ab}(k_s; t_1, t_2)
\sim \frac{k_{1234}+E_{k_s}}{E_{k_s} k_{1234} (k_{12}+E_s) (k_{34}+E_{k_s})} 
\eeq
where $a,b= \pm 1$,  and we defined $k_{1234} \equiv k_1+k_2+k_3+k_4$.
No resonances arise in the physical region.
\end{itemize}
The Schwinger–Keldysh formalism of eq.\eq{SKZ} introduced an artificial scalar $\phi_-$ that propagates backwards in time
and that does not interact with $\phi_+$ scalars.
In this formalism, {\em a ghost is a $\phi_-$ scalar that interacts with normal $\phi_+$ scalars}.
More precisely, a ghost $\phi$ scalar with opposite arrow of time is described, in the Schwinger–Keldysh formalism
with doubled field content, by reversing the roles of $\phi_\pm$  into $\phi_\mp$.
Following the Lee-Wick approach we compute amplitudes and correlators among normal scalars, mediated by a ghost mediator.
These are obtained by using for the ghost propagator
\beq  \label{eq:ghostProp}
-D_{-a, -b}(k_s; t_1, t_2)=-D_{ab}^*(k_s; t_1, t_2) \qquad  \hbox{rather than}\qquad D_{ab}(k_s; t_1,t_2).\eeq
The overall minus sign simply arises from the opposite sign action $S_-$.
The flipped arrow of time (corresponding to $-a$ and $-b$)
arises after resumming a negative decay width into the propagator, because
a small width is equivalent to an opposite $i\epsilon$ prescription. 
The complex conjugation identity means that, up the overall sign,
a ghost gives the same effects as a normal particle with $u_k(t)$ replaced by $u_k^*(t)$.
In flat space, this implies that the time-reversed ghost SK propagators are equal the SK propagators of a normal scalar flipping the sign of its energy, $E_{k_s}\to - E_{k_s}$.
So, the ghost-mediated amplitudes and correlators are given by the expressions mediated by a normal scalar, eq.\eq{flatcor}, 
replacing $E_{k_s} - i \epsilon\to - E_{k_s} -i \epsilon$.
As a consequence, a non-standard resonant enhancement appears in correlators.

\medskip

So far, we considered flat space. 
Field correlators are measured at the cosmo-collider, in a de Sitter inflationary background.

\section{What is a ghost in de Sitter?}\label{sec:dS}
We now consider a cosmological background, that provides one extra arrow of time, from cosmological expansion.
We approximate inflation as de Sitter, with metric $ds^2 = a^2 (d\tau^2 - d\vec{x}^2)$ where the 
 scale factor is $a=e^{H t}=-1/H\tau$ in terms of conformal time $\tau$.
The usual expanding de Sitter is described by the Poincar\'e branch with $-\infty<\tau<0$.
The branch $0<\tau<\infty$ describes a contracting de Sitter, irrelevant for inflation, but useful for our later discussion.

\subsection{Normal scalar in de Sitter}
The quantisation of a normal real scalar field $\phi(x)$ on a de Sitter background is a subtle standard topic, that we here summarise.
We assume the Lorentz-invariant action
\beq S_+=\int d^4 x \, a^3 \left[\frac{\dot\phi^2}{2} - \frac{(\nabla\phi)^2}{2a^2}-(m^2 + \xi R) \frac{\phi^2}{2}\right]\eeq
where $R=-12 H^2$ is the curvature.
Switching to conformal time $\tau$ and rescaling the field as $\varphi=a \phi$, the action takes the form
\beq S_+ = \int d^3x\,d\tau\left[ \frac{\dot\varphi^2}{2} - \frac{(\nabla\varphi)^2}{2}- M^2 \frac{\varphi^2}{2}\right]\eeq
where the mass $M$ depends on $\tau$
\beq M^2 = (m^2 + \xi R)a^2 - \frac{a''}{a} =\left[\frac{m^2}{H^2} - 12 \bigg(\xi + \frac16\bigg)\right]\frac{1}{\tau^2}.
\eeq
The mode functions $v_k = a u_k$ of the rescaled $\varphi$  field satisfy
\beq \frac{d^2 v_k}{d\tau^2} + \omega^2 v_k = 0,\qquad \omega^2 = k^2 + M^2 = k^2 + \frac{\mu^2+1/4}{\tau^2},\qquad
\mu = \sqrt{\frac{m^2}{H^2}-12\xi -\frac94}.
\eeq
One solution to the mode equations corresponds to the Bunch-Davies vacuum,
\beq u_k^{\rm BD}(\tau) =-i  e^{-\pi \mu/2+i \pi/4} H \frac{\sqrt{\pi}}{2} (-\tau)^{3/2} H^{(1)}_{i\mu}(-k\tau),\eeq
that  reduces to the flat-space vacuum at early times,
$v_k^{\rm BD}(\tau)  \stackrel{\tau\to-\infty}{\to} -i{e^{-i k\tau}}/{\sqrt{2k}}$.
A second independent solution to the mode equations is $u_k^{{\rm BD}*}(t)$. 
For large enough scalar mass, the parameter $\mu$ is real and modes keep oscillating after horizon exit, such that
cosmo-collider signals have oscillatory shape.
At late times
\beq  \label{eq:uknorm}
u_k^{\rm BD}(\tau) \stackrel{\tau\to 0}{\to}  \frac{(1-i)\sqrt{\pi/2} e^{-\pi\mu/2}(1 + \coth\pi\mu) }{2\Gamma(1+i \mu)}(-\tau)^{3/2}  H_{\rm infl} \bigg[
\bigg(-\frac{k\tau}{2}\bigg)^{i\mu} + r\bigg(-\frac{k\tau}{2}\bigg)^{-i\mu}  \bigg].\eeq
The first term in eq.\eq{uknorm} corresponds to the initial-state particle. 
The second term suppressed by 
\beq
r = \frac{\mu \Gamma(i\mu)^2}{\pi (1+\coth\pi\mu)}\stackrel{\mu\gg 1}{\simeq} e^{-\pi \mu - 2 i \mu (1-\ln\mu)}
\eeq
describes de Sitter particle production:
a thermal-like bath gets populated when a momentum $k$ is redishfted down to $H$.
The Schwinger–Keldysh propagators are again given by eq.\eq{D++}, inserting now the dS mode functions $u_k^{\rm BD}(\tau)$.
In particular, the Wightman functions $D_{\pm\mp}$ 
receive a contribution Boltzmann-suppressed by the de Sitter temperature, $e^{-2\pi \mu}\sim e^{-m/T_{\rm dS}}$ with $T_{\rm dS}=H/2\pi$.

\begin{figure}[t]
$$\includegraphics[width=0.7\textwidth]{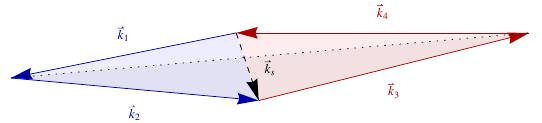}$$
\begin{center}
\caption{\em The momentum configuration of a 4-point correlator.\label{fig:k1234}}
\end{center}
\end{figure}

\medskip

We are interested in cosmo-collider effects, such as the bi-spectrum $\med{\zeta_{k_1}\zeta_{k_2}\zeta_{k_3}}$
and the tri-spectrum $A=\med{\zeta_{k_1} \zeta_{k_2} \zeta_{k_3} \zeta_{k_4}}$ evaluated at inflation end $\tau=0$.
We assume they are mediated by the virtual exchange of one $\phi$ particle produced at $\tau_1$ and absorbed at $\tau_2$.
Then, these correlators are given by products of couplings, 
external $\zeta$ propagators,  
and one internal $\phi$ propagator $D_{ab}(k_s;\tau_1,\tau_2)$.
One needs to integrate over $\tau_{1,2}$ and sum over the SK $a,b=\pm 1$ indices~\cite{1703.10166}.
The propagators of the external massless inflatons are proportional to $e^{\pm i k_i \tau_i}$ factors,
so that the final result can be conveniently decomposed in terms of `seed' integrals  similar to eq.\eq{flatcor}.
The normal seed integrals, denoted as $\overrightarrow{\cal I}$ to emphasize the normal arrow of time, are
\beq \label{eq:dScor}
\overrightarrow{\cal I}_{ab}^{p_1p_2}(r_1, r_2)\equiv -ab \frac{k_s^{5+p_1+p_2}}{H^2}
 \int_{-\infty}^0 d\tau_1 \, d\tau_2 \, (-\tau_1)^{p_1} (-\tau_2)^{p_2} \, e^{i a k_{12} \tau_1+i b k_{34} \tau_2} 
D_{ab}(k_s; \tau_1,\tau_2)
\eeq
where, again,
$k_i \equiv |\vec{k}_i|$, $k_{12}\equiv k_1+k_2$, $k_{34}\equiv k_3+k_4$ and $k_s\equiv |\vec{k}_1+\vec{k}_2| =  |\vec{k}_3+\vec{k}_4|$, see fig.\fig{k1234}.
Different powers of $p_{1,2}$ correspond to different terms in the external wave-functions and to different interactions.
Plausible interactions lead to $-4 \lesssim p_i \lesssim 0$, as summarised in the appendix.
The power of $k_s$ makes the seed integrals dimension-less  and dependent only on the ratios
\begin{align}
r_1\equiv \frac{k_s}{k_{12}},\qquad r_2\equiv \frac{k_s}{k_{34}}.
\end{align}
In flat space a $2\to 2$ scattering has a more general kinematical dependence than a $1\to 2$ decay.
Similarly, the tri-spectrum has a more general kinematical dependence than the bi-spectrum.
This allows to decompose the  tri-spectrum $A$
 as the sum of three terms with different physical origin (see e.g.~\cite{2112.03448})
\beq A = A_{\rm EFT} + A_{\rm L} + A_{\rm NL}.\eeq
\begin{enumerate}[a)]
\item The first {\em Effective Field Theory}  term is an analytic function of $r_1$ and of $r_2$.
It is equivalent to a local interaction term, similarly to how in flat space 
the virtual exchange of a heavy particle can be approximated as $1/(k^2 - m^2) \simeq - 1/m^2 + \cdots$
in terms of local EFT operators.

\item The second {\em local} term is non-analytic in $r_2/r_1=k_{12}/k_{34}$ and analytic in $r_1 r_2 = k_s^2/k_{12}k_{34}$,
such that, if transformed to coordinate space, it gives $\delta$ functions of position.
It corresponds to virtual  off-shell exchange, such as $1/(k^2 - m^2) \simeq 1/k^2 + \cdots$.

\item The third {\em non-local} term is non-analytic in $r_1 r_2$.
It comes from virtual resonant exchange, and is Boltzmann suppressed by $e^{-\pi\mu}$ factors,
as particle production arises from de Sitter.
\end{enumerate}
The difference between local and non-local is lost in the bi-spectrum, as it can be obtained as
$\lim_{k_4\to 0} \med{\zeta_{k_1} \zeta_{k_2} \zeta_{k_3}\zeta_{k_4}} \approx  \med{\zeta_{k_1} \zeta_{k_2} \zeta_{k_3}} \med{\zeta_0}$ 
in the limit $r_2\to 1$.
The difference between the three terms is lost in the power spectrum $\med{\zeta_k\zeta_{-k}}$, obtained in the limit $r_{1,2}\to 1$.

\smallskip

Cosmo-collider oscillatory signals arise in the squeezed limit $r_1\ll r_2\ll 1$.
Simple analytic expressions can be written in the limit of large $\mu$, $m \gg H$, showing that local and non-local oscillations
have the same Boltzmann-suppressed amplitude (see e.g.~\cite{2112.03448})
\beq \label{eq:Anormal}
\sum_{ab} \overrightarrow{\cal I}_{ab}^{p_1 p_2} \simeq -2\pi e^{-\pi\mu} r_1^{p_1+5/2} r_2^{p_2+5/2} \mu^{3+p_1+p_2} \bigg[ 
\underbrace{ \cos\bigg(\mu \ln \frac{r_1 r_2}{4} \bigg)}_{\hbox{\footnotesize non-local}} + 
\underbrace{\sin\bigg(\mu\ln \frac{r_1}{r_2}- \pi p_2 }_{\hbox{\footnotesize local}}\bigg) \bigg].\eeq
The Boltzmann suppression $e^{-\pi\mu}$ arises in a non-trivial way.
The integrand $u_k^{\rm BD}$ starts Boltzmann unsuppressed at $\tau\to -\infty$, and later acquires an extra 
Boltzmann suppressed contribution from de Sitter particle production.
The seed integrals over $\tau_{1,2}$ are Boltzmann suppressed because  Boltzmann unsuppressed terms in the internal propagator
oscillate in the same direction as the external propagators $e^{i a k_{12} \tau_1+i b k_{34} \tau_2} $,
so that the phase never gets stationary.

Our goal is exploring how this form, dictated by causality, changes if the mediator is a ghost with an opposite arrow of time.

\begin{figure}[t]
$$\includegraphics[width=0.435\textwidth]{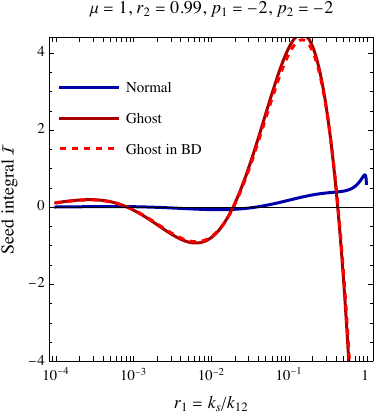}\qquad
\includegraphics[width=0.45\textwidth]{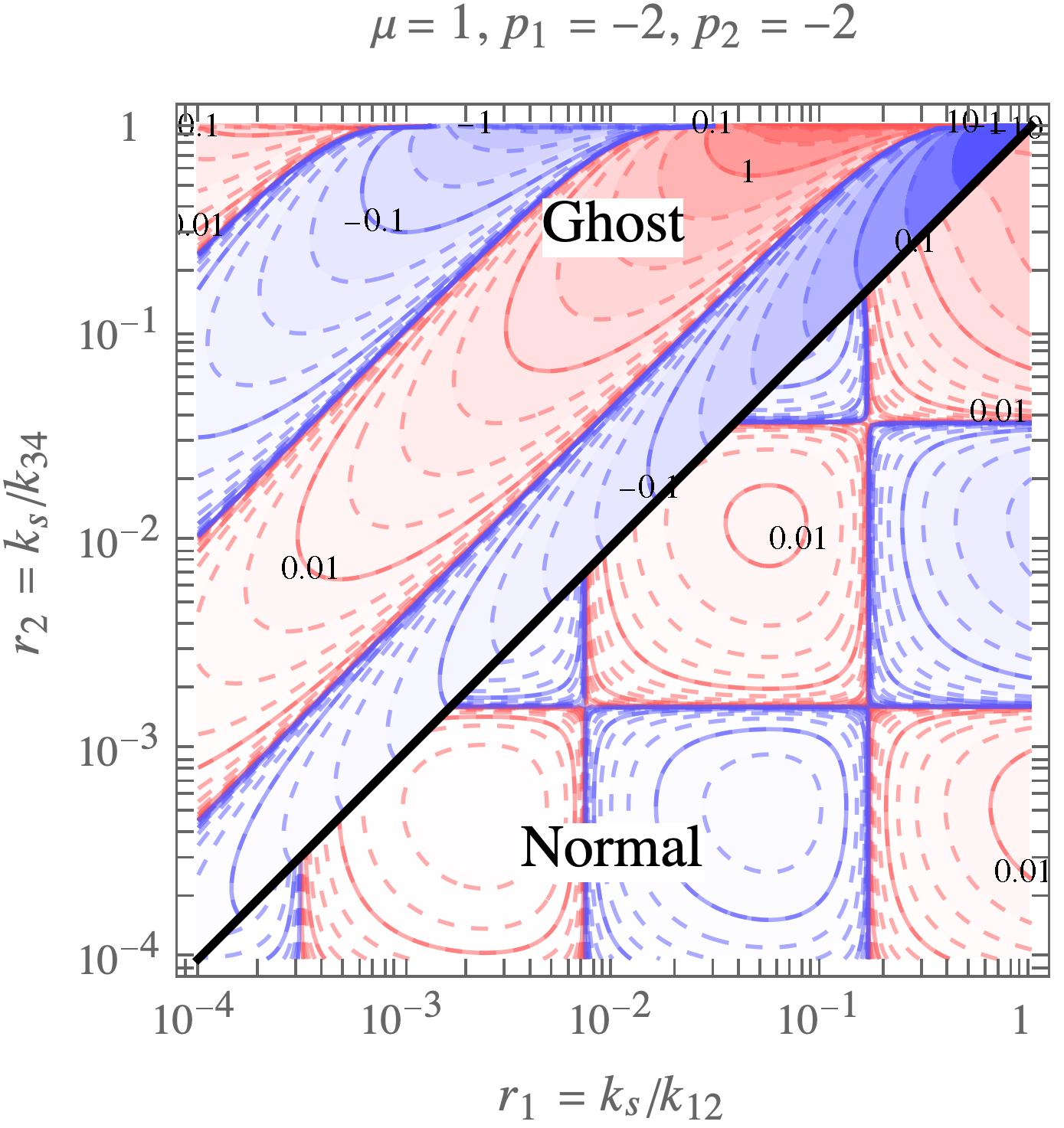}$$
\begin{center}
\caption{\em Sample seed functions ${\cal I}(r_1, r_2)$ for a normal particle and for a ghost.
\label{fig:SeedNormalGhostr1}
The left panel shows the result at fixed $r_2$.
The right panel displays the ghost $\overleftarrow{\cal I}(r_1,r_2)$ in the upper side, 
and the normal $\overrightarrow{\cal I}(r_2, r_1)$ in the lower side.
Regions of positive (negative) ${\cal I}$ are shaded in red (blue).
The different shapes reflect the absence of non-local effects in the ghost case, 
understood analytically in the in the $\mu\gg1 $ limit, eq.\eq{Ighost}.
The ghost seed integral is logarithmically divergent at $r_2\to 1$.}
\end{center}
\end{figure}

\subsection{Ghost in de Sitter}
We next address the quantisation of a ghost on a de Sitter background.
As in the case of a normal particle, we start assuming $|\Gamma |\ll m \sim H$ and
compute in the limit where the ghost life-time is much {\em longer} than the Hubble time, $|\Gamma|\ll H$.
Given that the ghost life-time is negative, $|\Gamma|\to 0$ is a subtle limit.
We identify two main ghost effects.

\subsubsection*{Time-reversed propagator}
A first ghost effect is that the ghost SK propagators take the form $-D_{-a,-b}$, as already discussed in flat space, eq.\eq{ghostProp}.
The overall minus sign leads to wrong-sign non-Gaussianities.
The opposite arrow of time  leads to more interesting effects.
The ghost seed integrals, denoted as $\overleftarrow{\cal I}$ to emphasize the reversed arrow of time, are
\begin{align} \label{eq:dScorghost}
\overleftarrow{\cal I}_{ab}^{p_1p_2}(r_1,r_2)& \equiv +ab \frac{k_s^{5+p_1+p_2}}{H^2}
 \int_{-\infty}^0 d\tau_1 \, d\tau_2 \, (-\tau_1)^{p_1} (-\tau_2)^{p_2} \, e^{i a k_{12} \tau_1+i b k_{34} \tau_2} 
D_{-a,-b}(k_s; \tau_1,\tau_2)
\end{align}
and contain a Boltzmann unsuppressed local term in the limit $\mu\gg 1$, $r_1\ll r_2\ll 1$:
\beq \label{eq:Ighostnaive}
\sum_{ab} \overleftarrow{\cal I}_{ab}^{p_1p_2}\simeq 2\pi  r_1^{p_1+5/2} r_2^{p_2+5/2} \mu^{3+p_1+p_2} \cos\bigg(\mu\ln \frac{r_1}{r_2}\bigg) + {\cal O}(e^{-\pi \mu}) .\eeq
So far we kept the $u_k^{\rm BD}$ mode functions.
Then, omitted terms suppressed by $e^{-\pi \mu}$ would contain a non-local oscillation.

\subsubsection*{Time-reversed mode functions}
 A second  ghost effect is that the ghost  propagators must plausibly be computed assuming mode
functions $u_k^{\rm ghost}(\tau)$ different from the Bunch-Davies mode functions $u_k^{\rm BD}(\tau)$.
Bunch-Davies assume the vacuum state at early times $\tau\to - \infty$, and obtain Boltzmann-suppressed particle production at late times $\tau\to 0$,
see eq.\eq{uknorm}.
This term must be subtracted, following the interpretation of a ghost as having a reversed arrow of time,
such that the ghost is in its vacuum state at late times $\tau\to 0$:
\beq \label{eq:ukghost}
u_k^{\rm ghost} =  \frac{u_k^{\rm BD} - i e^{-\pi\mu} u_k^{*\rm BD}}{\sqrt{1-e^{-2\pi\mu}}}=\frac{1-i}{2 \sqrt{\sinh\pi\mu}} (-\tau)^{3/2}H \sqrt{\pi } J_{i\mu}(-k\tau).\eeq
The effect of this subtraction alone is removing the non-local oscillation from eq.\eq{Anormal}.
This mode function is equivalent to setting an initial-time condition to a normal scalar on the time-reversed $\tau\to - \tau$ contracting branch of de Sitter.
We considered a massive enough scalar, $\mu\ge 0$, whose fluctuations are not frozen outside the horizon.

\subsection{Correlators with time-reversed propagator and mode functions}
We arrive at the final result by combining both effects: 
i) time-reversed ghost propagator with flipped sign, $-D_{-a-b} = - D_{ab}^*$; ii) time-reversed ghost mode functions as in eq.\eq{ukghost}.
The second effect adds
to eq.\eq{Ighostnaive} a Boltzmann-suppressed local term. 
In the limit $\mu\gg1$ and $r_1 \ll r_2 \ll 1$  the ghost seed integrals are given by
\beq \label{eq:Ighost}
\sum_{ab} \overleftarrow{\cal I}_{ab}^{p_1p_2} \simeq 2\pi  r_1^{p_1+5/2} r_2^{p_2+5/2} \mu^{3+p_1+p_2} \bigg[
\cos\bigg(\mu\ln \frac{r_1}{r_2}\bigg) + e^{-\pi\mu} \sin\bigg(\mu\ln\frac{r_1}{r_2}+\pi p_1\bigg)+{\cal O}(e^{-2\pi\mu})\bigg] .\eeq
The lack of a non-local oscillation persists expanding up to higher orders.

In appendix~\ref{sec:A} we present analytic results for
the seed integrals ${\cal I}$ computed assuming the most general mode function $u_k(\tau)$, parameterised as 
\beq  \label{eq:C+-} u_k = e^{i\theta} [C_+ u_k^{\rm BD} + C_-  e^{i\phi} u_k^{*\rm BD}].\eeq
This is useful, because time inversion acts as complex conjugation, such that
the ghost seed integrals $\overleftarrow{\cal I}$ equal the corresponding normal  $\overrightarrow{\cal I}$ seed integrals provided that $u_k$
is replaced by its complex conjugate, namely
\beq  \label{eq:Cswap}\overleftarrow{\cal I}(u_k) =-\overrightarrow{\cal I}(u_k^*). \eeq
Tree-level exchange of one ghost is thereby equivalent to tree-level exchange of a normal particle
in a mostly-antiparticle excited state.
This explains the lack of Boltzmann suppression:
from the point of view of normal particles, a ghost in its vacuum state contains excitations.
A related computation was performed in the context of $\alpha$ vacua~\cite{2309.05244}.

\medskip

Two qualitatively novel features of ghost seed integrals appear surprising.
First, the lack of a Boltzmann suppression makes the cosmo-collider potentially sensitive to ghosts with heavy mass $m\gg H$.
Second, a divergence $\sim 1/(1-r_2)^{2+p_2}$ appears in the bispectrum limit $r_2 \to 1$ of the ghost seed integrals.
Indeed, in this limit the ghost goes on-shell, $k_s\simeq  k_{34}$, such that the integral over $\tau_2$ acquires a stationary phase.
The seed integral in the critical log-divergent case $p_2 = -2$ is plotted in fig.\fig{SeedNormalGhostr1}.
The fact that divergences appear for  $p_2\ge -2$ signals that they arise from early times $H \tau_{2}\ll -1$.
Similar divergences $\sim 1/(1-r_1)^{2+p_1}$ appear in the power spectrum limit $r_1\to 1$.

\medskip

These two surprising enhancements are limited by a physical effect neglected so far.
Usual cosmo-collider computations assume a negligible decay rate $\Gamma \ll H$ of normal particles,
because the cosmo-collider can probe them up to mass $m \sim H$, to avoid the Boltzmann suppression.
Normal particles with perturbative couplings $g$ have a small decay rate $\Gamma \sim g^2 m/4\pi  \ll m $.
Detectable cosmo-collider effects need sizeable couplings, so one expects that $|\Gamma|$ is not much smaller than $m$.

\smallskip

In the ghost case, the previously neglected decay plays a more significant role.
In flat space, ghost-induced acausalities remain microscopic due to the decay factor $e^{-\Gamma t}$.
An analogous mechanism operates for ghost cosmo-collider acausalities.
The preceding results, derived in the $|\Gamma|\to 0$ limit, therefore remain valid only if
\beq |\Gamma| \ll H.\eeq
A decay rate comparable to the inflationary Hubble rate $H$ leads to novel effects.
Computing such loop-level effects is technically challenging and subtle even in the normal case~\cite{2407.08581}.
For ghosts, the situation is even more intricate: 
a ghost decays before it is produced, leading to phenomena that get enhanced at low momenta due to the inflationary redshift.
A $k_s$-dependent decay-like rate $\Gamma$ can be precisely defined in a de Sitter background~\cite{astro-ph/0406287}.
At early times, one may approximate the decay rate $\Gamma<0$ as constant.
Then, the additional decay factor 
$e^{-\Gamma t} = (-H \tau)^{\Gamma/H}$ provides, at early times, a shift $\delta p_{1,2} = \Gamma/H < 0$ in the dimension $p_2$ of the interaction.
So the divergence at $r_2\to 1$  is regulated only above a critical value of $|\Gamma|/H \gtrsim 2+p_2$, 
depending on the dimension of the ghost interaction operator.

For smaller values of $|\Gamma|/H$, a different physical  cut-off becomes relevant: 
the initial conformal time, approximated as $\tau_0 = -\infty $ in eq.\eq{dScorghost}, gets replaced by
the unknown onset of inflation, and perhaps even by the beginning of time itself.
This last consideration brings us uncomfortably close to the border between physics and speculation, so we conclude here.


\section{Conclusions}\label{concl}
Normal particles carry an intrinsic arrow of causality.
Lee-Wick ghosts, in contrast, are hypothetical particles with a reversed arrow of causality.
Theories where the two different kinds of particles interact feature acausalities on microscopic time-scales
of the order of the {\em negative} ghost life-time.

Tree-level exchange of a  ghost provides striking collider signals~\cite{0805.2156}.
Tree-level exchange of a normal scalar is the most plausible hope of a detectable cosmological collider signal.
We  studied the characteristic cosmo-collider signals that arise exchanging a Lee-Wick ghost.
We argued that a ghost propagating in the de Sitter inflationary background (which itself possesses a preferred arrow of time)
is described by 
\begin{enumerate}[a)]
\item replacing the usual Schwinger–Keldysh propagator $D_{ab}$ with minus its time-reversed counterpart, $-D_{-a,-b}$, and
\item substituting the usual
Bunch-Davies mode functions (which describe a field in the vacuum state at early times)
with the corresponding vacuum solution at late times.
\end{enumerate}
We identified the following characteristic cosmological-collider signatures arising from tree-level virtual exchange of a Lee-Wick ghost:
\begin{enumerate}
\item A ghost mediates non-Gaussianities with `wrong' overall sign that generically violates causality bounds,
and provides motivated specific predictions for the detailed signal.

\item A ghost mediates cosmo-collider oscillations of local type only, testable measuring the tri-spectrum.
In contrast, a normal scalar mediates both local and non-local oscillations, with comparable amplitudes (equal in the $\mu\gg 1$ limit).

\item Unlike ordinary fields, ghost exchange evades the usual Boltzmann suppression of cosmo-collider signals.

\item However, the possibility of detecting ghosts with mass $m\gg H$ is constrained by their finite (negative) decay rate.
 When $|\Gamma|\gtrsim H$, ghost-induced acausalities are suppressed due to a reduction, at early times, 
 of the effective scaling dimension of the interaction operator.
 
\item Depending on the dimensionality of the interaction, both the bispectrum and the power spectrum can exhibit infrared enhancements in the limits $k_i \to 0$.
\end{enumerate}
From a technical point of view, we computed the cosmological–collider seed integrals for a generic mode function, 
noting that a ghost prepared in its vacuum state appears to ordinary particles as a normal scalar field in a particular excited state.
Consequently, some of the signal features discussed above arise in more general scenarios involving non-Bunch-Davies initial conditions.

It would be interesting to investigate whether the massive spin-2 graviton ghost predicted by quadratic gravity 
(including conformal gravity) induces acausal effects that modify the minimal gravitational floor of primordial non-Gaussianities~\cite{Maldacena}.

 \small \paragraph{Acknowledgements.}  We thank Anish Ghoshal, Chat GPT, Grok, Zhehan Qin, Guilherme Pimentel,
 Alberto Salvio, Zhong-Zhi Xianyu and Yuhang Zhu for helpful discussion. S.A.\ is supported by the Japan Science and Technology Agency (JST) as part of Adopting Sustainable Partnerships for Innovative Research Ecosystem (ASPIRE),  grant JPMJAP2318.

\appendix\footnotesize

\section{Seed integrals for generic mode functions}\label{sec:A}
Computing cosmo-collider seed integrals is non trivial, both numerically and analytically.
We here study how ghost seed integrals can be reduced to normal seed integrals.

In flat space, reversing the arrow of time was equivalent to flipping the sign of the energy $E_{k_s}$.
However, de Sitter propagators are written in terms of momentum $k_s$.
Energy is not used as the de Sitter background violates its conservation by an amount $\Delta E_{k_s}\sim H$.
Eq.\eq{dScorghost} implies that ghost seed integrals can be expressed in terms
of normal seed integrals with negative $r_{1}=k_s/k_{12}$ and $r_2=k_s/k_{34}$, as
\beq\label{eq:-r} \overleftarrow{\cal I}_{ab}^{p_1p_2}(r_1,r_2)=-\overrightarrow{\cal I}_{-a,-b}^{p_1p_2}(-r_1, -r_2).\eeq
However, eq.\eq{-r} is impractical because known expressions for seed integrals in de Sitter assume positive  $0\le r_1 \le r_2 \le 1$~\cite{2208.13790,2301.07047}.
A more convenient approach makes use of eq.\eq{Cswap}:
ghost seed integrals $\overleftarrow{\cal I}$ can be computed 
as normal particle seed integrals $\overrightarrow{\cal I}$ for non Bunch-Davies mode functions.
In order to use this mathematical   trick, we here compute the normal seed integrals
extending the Mellin Barnes computation in~\cite{2208.13790} to the most general mode function of eq.\eq{C+-},
$u_k = e^{i\theta} [C_+ u_k^{\rm BD} + C_-  e^{i\phi} u_k^{*\rm BD}]$.
The parameters $C_\pm,\phi,\theta$ are real, and the overall phase $\theta$ is irrelevant.
In this language, the ghost mode functions of eq.\eq{ukghost} correspond to $C_+ = 1/\sqrt{1-e^{-2\pi\mu}}$ and $C_- e^{i\phi}= -i e^{-\pi\mu}C_+$.
So, the duality in eq.\eq{Cswap} implies that ghost seed integrals are obtained as normal seed integrals 
 by exchanging $C_+ \leftrightarrow C_-$. Keeping $\theta$ general  one needs
\beq C_+ e^{i\theta}=i e^{-\pi\mu}C_-, \qquad C_- e^{i(\theta+\phi)}= 1/\sqrt{1-e^{-2\pi\mu}}\eeq
Furthermore, the `naive' ghost results of eq.\eq{Ighostnaive} are obtained from the normal seed integrals for $C_+=0$ and $C_-=1$.

\medskip

The opposite-signs seeds are given by the factorised product of two $\tau$ integrals:
\begin{align}
\label{eq:I+-}
 \overrightarrow{\mathcal{I}}_{ \pm \mp}^{p_1 p_2}=&\ \frac{e^{\mp i \frac{\pi}{2} ({p}_{1}-p_2)}}{4\pi} r_1^{5 / 2+p_1} r_2^{5 / 2+p_2}\\
\nonumber &\times \sum_{n_1=0}^{\infty}\left\{\frac{(-1)^{n_1}}{n_{1} !} \left(\frac{r_1}{2}\right)^{2n_1+i \mu} \left(C_+\pm ie^{\pm \pi \mu}e^{\mp i \phi}C_-\right)
\Gamma\left[-n_1-i\mu  \right] \Gamma\left[p_1+\frac{5}{2}+2n_1+i\mu\right]+(\mu \rightarrow-\mu)\right\}\\
\nonumber&\times \sum_{n_2=0}^{\infty}\left\{\frac{(-1)^{n_2}}{n_{2} !} \left(\frac{r_2}{2}\right)^{2n_2+i \mu}\left(C_+\mp ie^{\mp \pi \mu}e^{\pm i \phi}C_-\right) 
\Gamma\left[-n_2-i\mu \right] \Gamma\left[p_2+\frac{5}{2}+2n_2+i\mu\right]+(\mu \rightarrow-\mu)\right\},
\end{align}
The same-sign seeds lead to double integrals over $\tau_{1,2}$, that can be conveniently divided in two terms 
${\cal I}_{\pm\pm} = {\cal I}_{\pm\pm\rm F} +{\cal I}_{\pm\pm\rm TO} $
by splitting the propagators assuming $r_1<r_2$ as 
\begin{align}
\label{eq:FTO}
 D_{\pm \pm}\left(k ; \tau_1, \tau_2\right)=D_{\mp\pm}\left(k ; \tau_1, \tau_2\right)+\left[D_{\pm\mp}\left(k ; \tau_1, \tau_2\right)-D_{\mp\pm}\left(k ; \tau_1, \tau_2\right)\right] \theta\left(\tau_2-\tau_1\right) .
\end{align}
The first term again has a simple Factorised form 
\begin{align}
\label{eq:I++}
 \overrightarrow{\mathcal{I}} _{ \pm \pm{\rm{F}}}^{p_1 p_2}=&\ \frac{\mp i e^{\mp i \frac{\pi}{2} (p_1+p_2)}}{4\pi} r_1^{5 / 2+p_1} r_2^{5 / 2+p_2}\\
\nonumber &\times \sum_{n_1=0}^{\infty}\left\{\frac{(-1)^{n_1}}{n_{1} !} \left(\frac{r_1}{2}\right)^{2n_1+i \mu} (-e^{\pm \pi \mu}C_+\pm ie^{\pm i \phi}C_-)
\Gamma\left[-n_1-i\mu \right] \Gamma\left[p_1+\frac{5}{2}+2n_1+i\mu\right]+(\mu \rightarrow-\mu)\right\}\\
\nonumber&\times \sum_{n_2=0}^{\infty}\left\{\frac{(-1)^{n_2}}{n_{2} !} \left(\frac{r_2}{2}\right)^{2n_2+i \mu}(+C_+\pm i e^{\pm \pi \mu}e^{\mp i \phi}C_-)  
\Gamma\left[-n_2-i\mu \right] \Gamma\left[p_2+\frac{5}{2}+2n_2+i\mu\right]+(\mu \rightarrow-\mu)\right\}.
\end{align}
The sums above can be performed analytically in terms of hypergeometric functions ${F}$, 
which converge rapidly at $r_{1,2}\ll 1$.
In this regime, the terms with lowest $n_{1,2}=0$ give the leading effect at $\mu\gg 1$.
Conversely, the series may develop divergences at $r_{1,2}\to 1$ depending on $p_{1,2}$.
The opposite-sign integrals could, in principle, be recast as same-sign integrals with one negative $r_{1,2}$;
however, this reformulation is again not convenient in practice, as made clear by comparing eq.\eq{I++} with eq.\eq{I+-}.

The only double integral containing a TimeOrdering arises from the second term in eq.\eq{FTO} and is
\begin{align}
\nonumber \overrightarrow{\mathcal{I}}_{ \pm \pm {\rm{TO}}}^{p_1 p_2}=&\ \frac{ i e^{\mp i \frac{\pi}{2} (p_1+p_2)}}{2\pi} r_1^{5 +p_1+p_2}\sum_{n_1,n_2=0}^{\infty}\frac{(-1)^{n_{1}+n_2}}{n_{1} !n_{2} !}  \left\{\left(\frac{r_1}{2}\right)^{2n_{1}+2 n_2} \left(C_+^2-C_-^2\right)\sinh (\pi \mu) 
\Gamma\left[-n_1+i\mu \right] \right.\\
\nonumber&\left.\times \Gamma\left[-n_2-i\mu\right]  \mathrm{~\mathcal{F}}\left[\begin{array}{c|c}
p_2+\frac{5}{2} +2n_2+i\mu,p_{1}+p_2+2n_1+2n_2+5   \\
p_2+\frac{7}{2} +2n_2+i\mu&
\end{array}  -\frac{r_1}{r_2}\right]+(\mu \rightarrow-\mu)\right\},
\end{align}
where
\begin{align}
\mathrm{~\mathcal{F}}\left[\begin{array}{c|c}
a,b   \\
c&
\end{array}  x\right]\equiv   \frac{\Gamma[a]\Gamma[b]}{\Gamma[c]}\ {}_2F_1\left[\begin{array}{c|c}
a,b   \\
c&
\end{array}  x\right], 
\end{align}
and ${}_2F_1[\cdots]$ is the hypergeometric function, analytic in $x$.
The TimeOrdered term is thereby of EFT form.
Furthermore, it has a qualitatively different dependence on $r_{1,2}$ that makes it suppressed in the $r_1\ll r_2 \ll 1$ limit.
The hypergeometric series converges for $r_1 < r_2$, as we have assumed. 
The results for $r_2 < r_1$ can be obtained by interchanging $r_1 \leftrightarrow r_2$ and $p_1\leftrightarrow p_2$~\cite{2112.03448}.
The integrals are convergent at $\tau_{1,2}\to 0$ for $p_{1,2}> -5/2$.
A closed form equivalent to the double sum was found using the alternative bootstrap method~\cite{2301.07047}.

\medskip

Finally, we summarise how specific interactions lead to specific seed integrals.
If the external $\zeta$ is a massless scalar with conformal $\xi=-1/6$,
its mode function is $\propto \tau e^{-i k \tau} \propto H_{-1/2}^{(1)}(-k\tau)$.
Then, the non-derivative $c_3 a^4 \zeta^2 \phi$ interaction leads to a tri-spectrum proportional to $\sum_{a,b} {\cal I}_{ab}^{-2,-2}$~\cite{2208.13790}.
If $\zeta$ is a massless scalar with minimal coupling  $\xi=0$, its mode function is $\propto (1+i k\tau)e^{-ik\tau} \propto H_{3/2}^{(1)}(-k\tau)$.
The two different powers of $\tau$ imply that the final result can be written as a sum
of seed integrals with different values of $p_{1,2}$. 
The derivative interaction $g a^2 (d\zeta/d\tau)^2 \sigma$ leads to
$\med{\zeta_{k_1} \zeta_{k_2} \zeta_{k_3} \zeta_{k_4}}'=g^2 (\sum_{a,b} {\cal I}_{ab}^{0,0})/4k_1k_2k_3k_4 k_s^5$~\cite{2208.13790},
where the prime denotes that we factored out an overall $(2\pi)^4 \, \delta(\vec{k}_1+\vec{k}_2+\vec{k}_3+\vec{k}_4)$.
The non-derivative interaction leads to
\begin{eqnarray}
\med{\zeta_{k_1} \zeta_{k_2} \zeta_{k_3} \zeta_{k_4}}' \!\!\!\!\!&=& \!\!\!\!\!
\frac{c_3^2 H^2}{4 k_1^3 k_2^3 k_3^3 k_4^3}\sum_{a,b = \pm}\bigg[
\frac{k_1 k_2 k_3 k_4}{k_s} {\cal I}^{-2,-2}_{ab} 
- i (a k_{12}k_3 k_4  {\cal I}^{-3,-2}_{ab}    + b k_1 k_2 k_{34}  {\cal I}^{-2,-3}_{ab} )
- ab k_{12} k_{34}k_s {\cal I}^{-3,-3}_{ab} 
\cr && 
- k_s (k_3 k_4  {\cal I}^{-4,-2}_{ab}  + k_1 k_2 k_s {\cal I}^{-2,-4}_{ab}  )
+ik_s^2 (ak_{12}   {\cal I}^{-3,-4}_{ab}  + b k_{34}   {\cal I}^{-4,-3}_{ab} )
+ k_s^3{\cal I}^{-4,-4}_{ab} +\cdots
\bigg]
\end{eqnarray}
where $\cdots$ denote 5 permutations from crossed diagrams.
The effective  $\Lag \supset c_2 a^3 \zeta'\phi + c_3 a^4 \zeta^2\phi$ describing 
a `turning' in field space~\cite{0911.3380} can generate a dominant contribution to the bi-spectrum, given by
\begin{align}
\left\langle\zeta_{{k}_1}\zeta_{{k}_2}\zeta_{{k}_3}\right\rangle^{\prime}    = 
 \frac{c_2c_3H}{4k_1^3k_2^3 k_3^2} \sum_{{a}, {b}= \pm} 
\lim_{k_4\rightarrow 0}\left[
 k_1k_2 \mathcal{I}_{{ab}}^{-2,-2}
-i{a}k_{12} k_3 \mathcal{I}_{{ab}}^{-3,-2}
-k_3^2  \mathcal{I}_{{ab}}^{-4,-2}\right]+2 \text { permutations. }
\end{align}
In the above results, $\zeta$ can be either the inflaton or the curvature perturbation; their different normalization can be absorbed in the couplings.


\footnotesize

\begin{thebibliography}{nnn}\bibitem{0911.3380}
\article[0911.3380]{X. Chen, Y. Wang}{JCAP}{04}{027}{2010}
{\href{https://doi.org/10.1088/1475-7516/2010/04/027}{Quasi-Single Field Inflation and Non-Gaussianities}}.


\bibitem{1109.0292}
\article[1109.0292]{D. Baumann, D. Green}{Phys.Rev.D}{85}{103520}{2012}
{\href{https://doi.org/10.1103/PhysRevD.85.103520}{Signatures of Supersymmetry from the Early Universe}}.


\bibitem{1211.1624}
\article[1211.1624]{T. Noumi, M. Yamaguchi, D. Yokoyama}{JHEP}{06}{051}{2013}
{\href{https://doi.org/10.1007/JHEP06(2013)051}{Effective field theory approach to quasi-single field inflation and effects of heavy fields}}.


\bibitem{1503.08043}
\heparticle[1503.08043]{N. Arkani-Hamed, J. Maldacena}{Cosmological Collider Physics}.


\bibitem{2205.01692}
\article[2205.01692]{Z. Qin, Z.-Z. Xianyu}{JHEP}{10}{192}{2022}
{\href{https://doi.org/10.1007/JHEP10(2022)192}{Phase information in cosmological collider signals}}.


\bibitem{1610.06597}
\article[1610.06597]{X. Chen, Y. Wang, Z.-Z. Xianyu}{Phys.Rev.Lett.}{118}{261302}{2017}
{\href{https://doi.org/10.1103/PhysRevLett.118.261302}{Standard Model Background of the Cosmological Collider}}.


\bibitem{1612.08122}
\article[1612.08122]{X. Chen, Y. Wang, Z.-Z. Xianyu}{JHEP}{04}{058}{2017}
{\href{https://doi.org/10.1007/JHEP04(2017)058}{Standard Model Mass Spectrum in Inflationary Universe}}.


\bibitem{1805.02656}
\article[1805.02656]{X. Chen, Y. Wang, Z.-Z. Xianyu}{JHEP}{09}{022}{2018}
{\href{https://doi.org/10.1007/JHEP09(2018)022}{Neutrino Signatures in Primordial Non-Gaussianities}}.


\bibitem{1811.11200}
\article[1811.11200]{S. Kumar, R. Sundrum}{JHEP}{04}{120}{2019}
{\href{https://doi.org/10.1007/JHEP04(2019)120}{Seeing Higher-Dimensional Grand Unification In Primordial Non-Gaussianities}}.


\bibitem{1907.07390}
\article[1907.07390]{S. Lu, Y. Wang, Z.-Z. Xianyu}{JHEP}{02}{011}{2020}
{\href{https://doi.org/10.1007/JHEP02(2020)011}{A Cosmological Higgs Collider}}.


\bibitem{2004.02887}
\article[2004.02887]{L.-T. Wang, Z.-Z. Xianyu}{JHEP}{11}{082}{2020}
{\href{https://doi.org/10.1007/JHEP11(2020)082}{Gauge Boson Signals at the Cosmological Collider}}.


\bibitem{2112.10793}
\article[2112.10793]{Y. Cui, Z.-Z. Xianyu}{Phys.Rev.Lett.}{129}{111301}{2022}
{\href{https://doi.org/10.1103/PhysRevLett.129.111301}{Probing Leptogenesis with the Cosmological Collider}}.


\bibitem{2203.06349}
\article[2203.06349]{X. Tong, Z.-Z. Xianyu}{JHEP}{10}{194}{2022}
{\href{https://doi.org/10.1007/JHEP10(2022)194}{Large spin-2 signals at the cosmological collider}}.


\bibitem{2408.07069}
\article[2408.07069]{S. Aoki, A. Ghoshal, A. Strumia}{JHEP}{11}{009}{2024}
{\href{https://doi.org/10.1007/JHEP11(2024)009}{Cosmological collider non-Gaussianity from multiple scalars and $R^{2}$ gravity}}.


\bibitem{1607.03735}
\article[1607.03735]{H. Lee, D. Baumann, G.L. Pimentel}{JHEP}{12}{040}{2016}
{\href{https://doi.org/10.1007/JHEP12(2016)040}{Non-Gaussianity as a Particle Detector}}.


\bibitem{2012.13667}
\article[2012.13667]{S. Aoki, M. Yamaguchi}{JHEP}{04}{127}{2021}
{\href{https://doi.org/10.1007/JHEP04(2021)127}{Disentangling mass spectra of multiple fields in cosmological collider}}.


\bibitem{2112.05710}
\article[2112.05710]{L. Pinol, S. Aoki, S. Renaux-Petel, M. Yamaguchi}{Phys.Rev.D}{107}{L021301}{2023}
{\href{https://doi.org/10.1103/PhysRevD.107.L021301}{Inflationary flavor oscillations and the cosmic spectroscopy}}.


\bibitem{2309.10849}
\article[2309.10849]{Z.-Z. Xianyu, J. Zang}{JHEP}{03}{070}{2024}
{\href{https://doi.org/10.1007/JHEP03(2024)070}{Inflation correlators with multiple massive exchanges}}.


\bibitem{2404.09547}
\article[2404.09547]{S. Aoki, L. Pinol, F. Sano, M. Yamaguchi, Y. Zhu}{JHEP}{09}{176}{2024}
{\href{https://doi.org/10.1007/JHEP09(2024)176}{Cosmological correlators with double massive exchanges: bootstrap equation and phenomenology}}.


\bibitem{1908.04170}
\article[1908.04170]{J.F. Donoghue, G. Menezes}{Phys.Rev.Lett.}{123}{171601}{2019}
{\href{https://doi.org/10.1103/PhysRevLett.123.171601}{Arrow of Causality and Quantum Gravity}}.


\bibitem{2003.09047}
\article[2003.09047]{J.F. Donoghue, G. Menezes}{Prog.Part.Nucl.Phys.}{115}{103812}{2020}
{\href{https://doi.org/10.1016/j.ppnp.2020.103812}{Quantum causality and the arrows of time and thermodynamics}}.


\bibitem{2502.04215}
\article[2502.04215]{L. Hui, A. Nicolis, A. Podo, S. Zhou}{JHEP}{07}{188}{2025}
{\href{https://doi.org/10.1007/JHEP07(2025)188}{Microcausality without Lorentz invariance}}.


\bibitem{2506.19198}
\heparticle[2506.19198]{C. de Rham, S. Jazayeri, A.J. Tolley}{Bispectrum Islands}.


\bibitem{0805.2156}
\article{T.D.~Lee, G.C.~Wick}{Nucl.\ Phys. B}{9}{209}{1969}{Negative Metric and the Unitarity of the S Matrix}.
\article{S. Coleman}{7th International School of Subnuclear Physics}{}{282}{1969}{Acausality}.
\article[0805.2156]{B. Grinstein, D. O'Connell, M.B. Wise}{Phys.Rev.D}{79}{105019}{2009}
{\href{https://doi.org/10.1103/PhysRevD.79.105019}{Causality as an emergent macroscopic phenomenon: The Lee-Wick ${\cal O}(N)$ model}}.


\bibitem{1512.01237}
\article[1512.01237]{A. Salvio, A. Strumia}{Eur.Phys.J.C}{76}{227}{2016}
{\href{https://doi.org/10.1140/epjc/s10052-016-4079-8}{Quantum mechanics of 4-derivative theories}}.


\bibitem{1709.04925}
\article[1709.04925]{A. Strumia}{MDPI Physics}{1}{17}{2019}
{\href{https://doi.org/10.3390/physics1010003}{Interpretation of quantum mechanics with indefinite norm}}.


\bibitem{1806.03605}
\article[1806.03605]{D. Anselmi, M. Piva}{JHEP}{11}{021}{2018}
{\href{https://doi.org/10.1007/JHEP11(2018)021}{Quantum Gravity, Fakeons And Microcausality}}.


\bibitem{1908.02416}
\article[1908.02416]{J.F. Donoghue, G. Menezes}{Phys.Rev.D}{100}{105006}{2019}
{\href{https://doi.org/10.1103/PhysRevD.100.105006}{Unitarity, stability and loops of unstable ghosts}}.


\bibitem{2004.00376}
\article[2004.00376]{P.D. Mannheim}{Int.J.Mod.Phys.D}{29}{2043009}{2020}
{\href{https://doi.org/10.1142/S0218271820430099}{Ghost problems from Pauli–Villars to fourth-order quantum gravity and their resolution}}.


\bibitem{2110.02246}
\article[2110.02246]{B. Holdom}{JHEP}{04}{133}{2022}
{\href{https://doi.org/10.1007/JHEP04(2022)133}{Photon-photon scattering from a UV-complete gravity QFT}}.


\bibitem{Stelle:1976gc}
\article{K.S. Stelle}{Phys.Rev.D}{16}{953}{1977}
{\href{https://doi.org/10.1103/PhysRevD.16.953}{Renormalization of Higher Derivative Quantum Gravity}}.


\bibitem{1703.10166}
\article[1703.10166]{X. Chen, Y. Wang, Z.-Z. Xianyu}{JCAP}{12}{006}{2017}
{\href{https://doi.org/10.1088/1475-7516/2017/12/006}{Schwinger-Keldysh Diagrammatics for Primordial Perturbations}}.


\bibitem{2308.14802}
\article[2308.14802]{Z. Qin, Z.-Z. Xianyu}{JHEP}{01}{168}{2024}
{\href{https://doi.org/10.1007/JHEP01(2024)168}{Nonanalyticity and on-shell factorization of inflation correlators at all loop orders}}.


\bibitem{2208.13790}
\article[2208.13790]{Z. Qin, Z.-Z. Xianyu}{JHEP}{04}{059}{2023}
{\href{https://doi.org/10.1007/JHEP04(2023)059}{Helical inflation correlators: partial Mellin-Barnes and bootstrap equations}}.


\bibitem{2112.03448}
\article[2112.03448]{X. Tong, Y. Wang, Y. Zhu}{JHEP}{03}{181}{2022}
{\href{https://doi.org/10.1007/JHEP03(2022)181}{Cutting rule for cosmological collider signals: a bulk evolution perspective}}.


\bibitem{2309.05244}
\article[2309.05244]{Y. Yin}{Phys.Rev.D}{109}{043535}{2024}
{\href{https://doi.org/10.1103/PhysRevD.109.043535}{Cosmological collider signal from non-Bunch-Davies initial states}}.


\bibitem{2407.08581}
\article[2407.08581]{T. Cohen, D. Green, Y. Huang}{Phys.Rev.D}{111}{103513}{2025}
{\href{https://doi.org/10.1103/PhysRevD.111.103513}{Operator origin of anomalous dimensions in de Sitter space}}.


\bibitem{astro-ph/0406287}
\article[astro-ph/0406287]{D. Boyanovsky, H.J. de Vega}{Phys.Rev.D}{70}{063508}{2004}
{\href{https://doi.org/10.1103/PhysRevD.70.063508}{Particle decay in inflationary cosmology}}.


\bibitem{2301.07047}
\article[2301.07047]{Z. Qin, Z.-Z. Xianyu}{JHEP}{07}{001}{2023}
{\href{https://doi.org/10.1007/JHEP07(2023)001}{Closed-form formulae for inflation correlators}}.

\bibitem{Maldacena}
\article[astro-ph/0210603]{J.M. Maldacena}{JHEP}{05}{013}{2003}
{\href{https://doi.org/10.1088/1126-6708/2003/05/013}{Non-Gaussian features of primordial fluctuations in single field inflationary models}}.


\end{thebibliography}
\end{document}